\documentclass{INTERSPEECH2023}

\interspeechcameraready

\usepackage{booktabs,multirow}
\title{A Novel Transfer Learning Method Utilizing Acoustic and Vibration Signals for Rotating Machinery Fault Diagnosis}
\name{Zhongliang Chen$^1$, Zhuofei Huang$^1$, Wenxiong Kang$^1$}
\address{
  $^1$South China University of Technology, China}
\email{202320116428@mail.scut.edu.cn, }

\begin{document}

\maketitle
 
\begin{abstract}
Fault diagnosis of rotating machinery plays a important role for the safety and stability of modern industrial systems. However, there is a distribution discrepancy between training data and data of real-world operation scenarios, which causing the decrease of performance of existing systems. This paper proposed a transfer learning based method utilizing acoustic and vibration signal to address this distribution discrepancy. We designed the acoustic and vibration feature fusion MAVgram to offer richer and more reliable information of faults, coordinating with a DNN-based classifier to obtain more effective diagnosis representation. The backbone was pre-trained and then fine-tuned to obtained excellent performance of the target task. Experimental results demonstrate the effectiveness of the proposed method, and achieved improved performance compared to STgram-MFN.
\end{abstract}

\section{Introduction}

Fault diagnosis of rotating machinery plays a important role for the safety and stability of modern industrial systems, which aims to precisely identify specific fault pattern of rotating machinery. Over the past decades, a significant amount of efforts have been dedicated to develop artificial intelligence (AI) for this task, including machine learning and deep learning methods. Regarding machine learning methods, i.e., supported vector machine (SVM), k-nearest neighbor (KNN), random forest (RF), and ANN have been widely used in fault diagnosis of rotating machinery. In terms of deep learning, most methods are based on neural network such as convolutional neural network (CNN), deep neural network (DNN), recurrent neural network (RNN) and stacked auto-encoder (SAE). These methods despite machine learning or deep learning, are based on the assumption that the training and testing data come from the same distribution. However, this assumption dose not hold in real-world operation scenarios as the data distribution can vary across the different operating conditions and different machines. This distribution discrepancy between datasets can significantly degrade the performance of the aforementioned AI systems.

In order to tackle this discrepancy, transfer learning (TL) has been applied in fault diagnosis of rotating machinery over the past few years. By transferring the knowledge obtained from one or more source tasks to the new target task, performance of deep networks on the target task is enhanced. In this manner, deep networks can be pre-trained using sufficient data of source domain and then fine-tuned utilizing a small amount of data of target domain to obtain excellent performance of target task\cite{pei2021rotating}. Thus, for fault diagnosis of rotating machinery, a lot of TF methods have been proposed, which mostly are vibration signal-based. For instance, a deep convolutional auto-encoding neural network was designed to fine-tune with a small number of labeled samples to address planetary gearbox fault diagnosis problem \cite{doi:10.1177/1748006X20964614}. In \cite{7961149}, higher accuracy and less training time can be achieved since they utilize a improved neural network to fine-tune. To realize high-precision and high-efficiency, a novel deep learning framework combining raw vibration signal of multiple sensors to input VGG19 was proposed \cite{Zhou2020MultisignalVN}. In terms of strong
feature extraction ability, fewer parameters and less training costs, an improved CNN consisting of multi-scale convolutional layer (MSC), a channel attention layer (CA), and an inception network structure (INS) was presented in \cite{Jiang_2022}. The extension of the application of Transformer architectures to rotating machinery fault diagnosis was first shown in \cite{pei2021rotating}, and the exceptional robustness and effectiveness of the proposed Transformer convolution network (TCN) was demonstrated. The abovementioned TL-based methods have obtained satisfactory performance in rotating machinery fault diagnosis. However, these methods only consider the vibration signal-based processing technique. This single-modal TL-based fault diagnosis often fails to take into account the complex nature of faults \cite{wang2021bearing}. Furthermore, relying solely on a single-modal can bring missed detections and vulnerability to external interferences. Consequently, these methods may not obtain a sufficiently reliable fault diagnosis. It is essential to adopt a multi-modal method to mitigate missed detections and enhance accuracy of fault diagnosis. Additionally, acoustic signal-based methods were developed gradually. The acoustic and vibration signal generated by rotating machinery often comprise the characteristics of fault related signals generated by machine components such as gears, bearings, and cam. Compared to vibration signal-based methods, sensors used for acquiring acoustic signal have relatively lower cost and faster measurement speed \cite{amarnath2013exploiting}. Thus, it is promising and feasible to fuse features of acoustic and vibration signal to conduct TL-based rotating machinery fault diagnosis.

\begin{figure*}[!t]
  \centering
  \includegraphics[width=\linewidth]{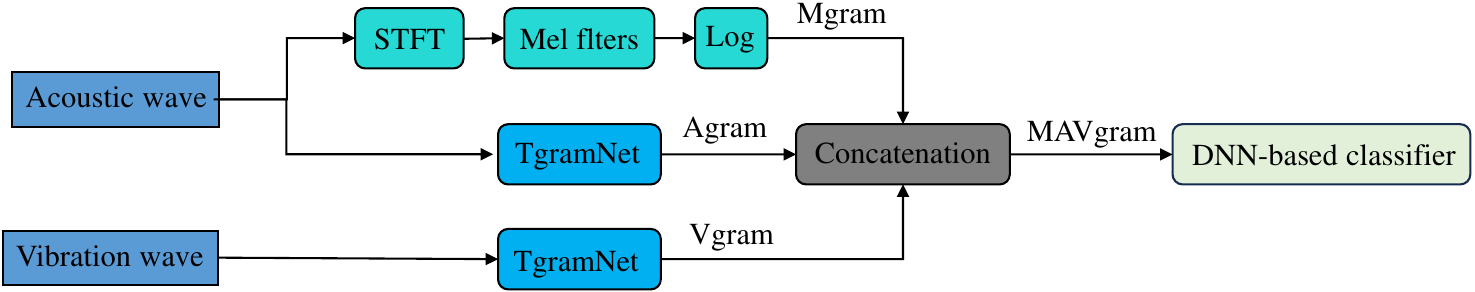}
  \caption{Framework of the proposed method.}
  \label{fig:backbone_structure}
\end{figure*}

In this study, we design a acoustic-vibration feature fusion MAVgram inspired by STgram-MFN \cite{liu2022anomalous} which achieved state-of-the-art performance on the Detection and Classification of Acoustic Scenes and Events (DCASE) Challenge 2020 Task 2 dataset as a outstanding anomalous sound detection (ASD). STgram is a spectral-temporal information fusion of acoustic signal, which consists of Log-Mel spectrogram (Sgram) and Tgram extracted by a CNN block TgramNet from raw acoustic signal. Tgram can compensate the important information of faults unavailable from Sgram, as Sgram filters out high-frequency components of acoustic signal causing ineffectiveness in distinguishing normal and anomalies \cite{zhang2023anomalous}. Based on the effectiveness of spectral-temporal information fusion, we followed STgram as the acoustic feature (MAgram) extracted from acoustic signal and applied TgramNet to extract vibration feature (Vgram) from vibration signal. Subsequently, we concatenate them together to compose MAVgram. MAVgram is then fed into a DNN-based classifier, i.e., MobileFaceNet (MFN), to learn the precise representations of different rotating machinery faults. To enhance robustness and discriminative capability of the backbone, speed perturbation is employed to augment the training data. Arcface \cite{deng2019arcface} is utilized as the loss function which helps enhance intra-class compactness and magnifying inter-class differences. Besides, in order to address the discrepancy between training data and data of real-world operation scenarios, we adopted TL technique including two steps: (i) pre-training the backbone on the source dataset; (ii) fine-tuning the backbone on the target dataset. The pre-training helps the backbone to acquire common knowledge on source dataset, and the fine-tuning adapts the backbone to the target domain. To the best of our knowledge, the proposed method is the first to utilizing acoustic-vibration signal in TL-based methods for rotating machinery fault diagnosis. Our experiments demonstrate that the proposed
method can obtain promising results with a very limited amount of data used for fine-tuning, and a ablation study was conducted for evaluating the effectiveness of the proposed acoustic-vibration feature fusion MAVgram.

\section{Proposed method}

\subsection{Extracting acoustic feature}

In this section, two main acoustic features are extracted: the audio temporal feature (Agram) and the log-Mel spectrogram feature (Mgram). Let $ x_{a} \in \mathbb R^{1 \times L}  $ represents the input acoustic signal with one channel and the length $ L $. Mgram $ F_{Mgram} \in \mathbb R^{M \times N}$ denotes the log-Mel spectrogram of $ x_{a} $, where $ M $ is the number of the Mel bins and $ N $ is the number of temporal frame. Mgram is constructed as following:
\begin{align}
 F_{Mgram} = log(\omega_{M}\cdot\Vert STFT(x_{a})\Vert^{2}) 
\end{align}
where $ STFT( \cdot ) $ represents short-time Fourier transform (STFT), $ \omega_{M} \in \mathbb R^{M \times B} $ denotes the Mel-filter weight used for the spectrogram which is derived from STFT and $ B $ is the number of frequency bins of the spectrogram.

Mgram mainly contains spectral feature of the input acoustic signal $ x_{a} $, which lacks the crucial temporal feature. To compensate the missing temporal information, a CNN module (TgramNet) is used for extracting the temporal feature of $ x_{a} $. TgramNet is mainly composed of a large kernel 1D convolution and three CNN blocks. Each CNN  block consists of a layer normalization, a leaky ReLU activation function, and a 1D convolution with a smaller kernel size. Thus, the resulting temporal feature Agram can be obtained:
\begin{align}
 F_{Agram} = TN(x_{a}) 
\end{align}
where $ TN( \cdot ) $ denotes the TgramNet, and the dimension of $ F_{Agram} \in \mathbb R^{M \times N} $ is the same as $ F_{Mgram} $. At this stage, temporal feature and spectral feature of the input acoustic signal have been extracted.

\subsection{Extracting vibration feature}

In order to complement valuable fault information utilizing vibration signal, temporal feature of the vibration signal is also extracted specially. Let $ x_{v} \in \mathbb R^{1 \times K}  $ represents the input vibration signal with one channel and the length $ K $. TgramNet  mentioned earlier is employed for extracting temporal feature of vibration signal (Vgram). Correspondingly, Vgram is :
\begin{align}
 F_{Vgram} = TN(x_{v}) 
\end{align}
where $ F_{Vgram} \in \mathbb R^{M \times N} $ has the same dimension as $ F_{Mgram} $. It's worth noting that the TgramNet here used for extracting Vgram and the previous TgramNet used for extracting Agram refer to two distinct models. For the sake of convenience in notation, we adopt a consistent expression $ TN( \cdot ) $ to denote TgramNet.
 
\subsection{Merging acoustic-vibration feature}

Subsequently, the acoustic-vibration fusion feature MAVgram $ F_{MAVgram}\in\mathbb R^{3 \times M \times N} $ is obtained adopting a simple mergence way by concatenating Mgram, Agram and Vgam sequentially:
\begin{align}
 F_{MAVgram} = Concat(F_{Mgram}, F_{Agram}, F_{Vgram}) 
\end{align}
where $ Concat(\cdot) $ represents the concatenation operation. 

\subsection{Pre-training the backbone on the source dataset}

In order to enhance the representativeness of the embeddings obtained from MAVgram to capture rich information about various faults, we adopt a DNN-based classifier for training. The overall framework of the proposed method is shown in Figure \ref{fig:backbone_structure}. Specially, we choose MFN as our baseline classifier to learn the fault information from MAVgram. Since the feature MAVgram fed into the classifier is extracted from both acoustic and vibration signal, the baseline (MAVgram-MFN) can learn richer and more reliable fault information compared to STgram-MFN. In addition, speed perturbation is utilized during training stage to enhance the backbone's discriminative capability and robustness. This data augmentation technique simulates virtual fault patterns to increase the diversity of data by modifying the speed of the original acoustic and vibration signal. For better discriminative capability to the various fault patterns, Arcface is utilized as it helps expand the distance between classes and reduce the distance within classes.

\subsection{Fine-tuning the backbone on the target dataset}

Since the fault patterns of the target task could be out-of-distribution, we need to fine-tune the pre-trained backbone on the target dataset. Fine-tuning helps adapt the backbone to the target task utilizing a limited amount of data, saving a lot of time and exempting the need of massive training data \cite{anbalagan2023foundational}. During the fine-tuning stage, the weights of TgramNet and the last full connected layer of MFN will be trained while other layers' weights are frozen to remain unchanged. The frozen weights maintain the outstanding ability of backbone and avoid overfitting on the target dataset. Let TgramNet and the last full connected layer of MFN being trained helps backbone be capable of classifying data with different sampling characteristics and adapt to target task precisely. After fine-tuning, the backbone can obtain exceptional performance on the target task.
\begin{table}[ht]
	\caption{Description of iFlytek and UO datasets}
 	\centering
	\label{tab:dataset}
	\resizebox{\linewidth}{!}{
	\begin{tabular}{cccc} 
		\toprule 
        Dataset & Fault pattern & Number of samples & Label category \\
        \midrule
        \multirow{5}{*}{iFlytek}  & normal  & 480 & 0 \\
        ~ & outer race fault & 480 & 1 \\
		~ & inner race fault & 480 & 2 \\
        ~ & rolling element fault & 480 & 3 \\
        ~ & cage fault & 480 & 4 \\
        \midrule
        \multirow{5}{*}{UO} & normal  & 380 & 0 \\
        ~ & outer race fault & 190 & 1 \\
        ~ & inner race fault & 190 & 2 \\
        ~ & ball fault & 190 & 3 \\
        ~ & cage fault & 190 & 4 \\
        \bottomrule 
	\end{tabular}}
\end{table}

\section{Experiments setup}

During the backbone training phase, we utilize the dataset provided by iFlytek for the rotating machinery fault diagnosis competition to train backbone weights. The dataset consists of synchronized acoustic and vibration data. The acoustic data was collected using microphones, with a sampling frequency of 48000 Hz and 16-bit quantization. The vibration data was collected using wired vibration acceleration sensors horizontally mounted on the bearing housing, with a sampling frequency of 5120 Hz. This dataset encompasses normal state and four distinct fault locations: outer race fault, inner race fault, rolling element fault, and cage fault. Each category includes acoustic and vibration data from 16 different operating conditions. We consider an additional dataset from University of Ottawa (UO) as target dataset to fine-tune the trained backbone \cite{sehri2023university}. UO dataset consists of acoustic and vibration data collected at a sampling rate of 42kHz under constant load and speed condition. It contains data belongs to one healthy state and four fault locations: outer race fault, inner race fault, ball fault, and cage fault. The data for healthy bearings, as well as bearings with inner, outer, and cage faults, was recorded under a consistent nominal load of 400 N. For ball fault data, no load was applied because ball faults developed naturally within a reasonable amount of time under degreased conditions when compared to the other faulty types. The description of iFlytek and UO datasets is shown in Table \ref{tab:dataset}.

For improving model performance, training speed and convergence effectiveness, min-max normalization is utilized during data preprocessing stage. The duration of per acoustic and vibration sample of iFlytek dataset and UO dataset is 4 seconds and 1 second, respectively. The number of frame is 376 samples for Mgram, and the number of Mel filter banks is 64. Hence, Mgram, Agram and Vgram have a dimension of $ {64 \times 376} $. The speed perturbation parameters were set as $n=3$ and $s=0.1$, where $n$ denotes the number of total speed categories and $s$ denotes the step of speed perturbation. The margin and scale parameters of ArcFace loss function are 0.7 and 30, respectively. The backbone model is trained and fine-tuned with 200 epochs and batch size 32. Adam optimizer is utilized for model training with a learning rate of 0.0005, and the cosine annealing strategy is employed for learning rate adjustment. The performance is evaluated by average accuracy of all fault patterns.

\begin{table}[ht]
	\caption{Performance comparison on different backbones (\%). }
 	\centering
	\label{tab:result1}
	\resizebox{\linewidth}{!}{
	\begin{tabular}{cccccc} 
		\toprule 
		\multirow{2}{*}{Method} & \multicolumn{5}{c}{Percentage of data for fine-tuning} \\
		\cmidrule(lr){2-6} 
		~ & 5\% &  10\% &  15\% &  20\% &  25\%  \\
		\midrule
		MAVgram-MFN &  \textbf{86.00} & \textbf{89.85} & \textbf{93.47} & \textbf{94.28} & \textbf{94.98} \\
		
		MAVgram-MobileNetV2 &  83.78 & 84.01 & 87.51 & 88.21 & 93.35 \\
		
		MAVgram-ResNet18 &  85.06 & 89.26 & 90.43 & 91.60 & 92.07 \\
		\bottomrule 
	\end{tabular}}
\end{table}

\section{Results}
Table \ref{tab:result1} shows the comparison of different backbones referring to MAVgram-MFN, MAVgram-MobileNetV2 and MAVgram-ResNet18. 
We fine-tuned the backbones on different percentages of data which is sampled from UO dataset to research the generalization performance of MAVgramNet. The data for testing which is fixed is composed of 75\% of UO dataset. Evidently MAVgram-MFN outperforms the other methods on each percentage of data for fine-tuning, and it can reach 0.9347 accuracy with only 15\% of data. This shows that the proposed method MAVgramNet can obtain high accuracy with a very limited amount of data used for fine-tuning.

For a better understanding of the impact of various input features, an ablation work is conducted in Table \ref{tab:ablation}. We adopted AVgram, STgram and MVgram as the input feature of MFN respectively. Table \ref{tab:ablation} shows that AVgram-MFN exhibits a significant decrease in accuracy compared to baseline on every percentage of data. The decrease of AVgram-MFN in accuracy is mainly due to the absence of spectral information which plays a critical role in acoustic feature. As for STgram-MFN and MVgram-MFN, their performance is comparable, showing a smaller accuracy on 10\% to 25\% of data. Especially, there is approximately a 10\% decrease in accuracy on 5\% of data. Compared to the above input feature alternatives, the proposed MAVgram is much more superior and effective for rotating machinery fault diagnosis as reported in Table \ref{tab:ablation}, especially for minimal data.
\begin{table}[ht]
	\caption{Ablation study on different input features (\%).}
	\label{tab:ablation}
	\centering
	\resizebox{\linewidth}{!}{
	\begin{tabular}{cccccc} 
		\toprule 
		\multirow{2}{*}{Method} & \multicolumn{5}{c}{Percentage of data for fine-tuning} \\
		\cmidrule(lr){2-6} 
		~ & 5\% &  10\% &  15\% &  20\% &  25\%  \\
		\midrule
		baseline & 86.00 & 89.85 & 93.47 & 94.28 & 94.98\\
		\midrule
		\midrule
		AVgram-MFN & 32.32 & 42.82 & 54.96 & 60.68 & 62.78\\
		STgram-MFN & 75.03 & 87.63 & 88.68 & 90.78 & 92.53\\
		MVgram-MFN & 74.33 & 88.91 & 89.38  & 90.43 & 91.95 \\
		\bottomrule 
	\end{tabular}}
\end{table}

Subsequently, we analyse the impact of $n$ and $s$ parameters of speed perturbation in Table \ref{tab:sp_result}. It is worth noting that we use $n=3$, $s=0.1$ in baseline. And we observe a slight drop in performance when only $s$ decreases since the discrepancy between practical and virtual class narrows. Especially, MAVgram-MFN reaches its best performance when $n=7$, $s=0.1$. However, it will cost much more time on backone training and fine-tuning stages as $n$ increases the amount of training data linearly.
\begin{table}[ht]
	\caption{Analysis of the impact of different parameters of speed perturbation (\%).}
	\label{tab:sp_result}
	\centering

	\begin{tabular}{cccccc} 
		\toprule 
		Method & $n$ & $s$ & Accuracy \\
		\midrule
		\multirow{9}{*}{MAVgram-MFN} & 3 & 0.1 & 94.98 \\
		~ & 3 & 0.05 & 92.77 \\
		~ & 3 & 0.025 & 88.80 \\
		~ & 5 & 0.1 & 93.70 \\
		~ & 5 & 0.05 & 90.90 \\
		~ & 5 & 0.025 & 94.05 \\
		~ & 7 & 0.1 & \textbf{95.68} \\
		~ & 7 & 0.05 & 93.12 \\
		~ & 7 & 0.025 & 94.17 \\
		\bottomrule 
	\end{tabular}
\end{table}

\section{Conclusion}
In this paper we proposed a novel transfer learning method utilizing acoustic-vibration signal for rotating machinery fault diagnosis , where vibration feature and acoustic feature are first extracted and then merged to input a DNN-based classifier for better diagosis performance. The proposed method applies pre-training and fine-tuing to excellently adapt to the target task using a limited amount of data, and speed perturbation is employed for enhancing robustness and discriminative capability of backbone. The experimental results demonstrate the effectiveness of the proposed method, and more feature fusion approaches will be explored in future work.

\nocite{*}
\bibliographystyle{IEEEtran}
\bibliography{ref}

\begin{thebibliography}{10}
\providecommand{\url}[1]{#1}
\csname url@samestyle\endcsname
\providecommand{\newblock}{\relax}
\providecommand{\bibinfo}[2]{#2}
\providecommand{\BIBentrySTDinterwordspacing}{\spaceskip=0pt\relax}
\providecommand{\BIBentryALTinterwordstretchfactor}{4}
\providecommand{\BIBentryALTinterwordspacing}{\spaceskip=\fontdimen2\font plus
\BIBentryALTinterwordstretchfactor\fontdimen3\font minus \fontdimen4\font\relax}
\providecommand{\BIBforeignlanguage}[2]{{%
\expandafter\ifx\csname l@#1\endcsname\relax
\typeout{** WARNING: IEEEtran.bst: No hyphenation pattern has been}%
\typeout{** loaded for the language `#1'. Using the pattern for}%
\typeout{** the default language instead.}%
\else
\language=\csname l@#1\endcsname
\fi
#2}}
\providecommand{\BIBdecl}{\relax}
\BIBdecl

\bibitem{pei2021rotating}
X.~Pei, X.~Zheng, and J.~Wu, ``Rotating machinery fault diagnosis through a transformer convolution network subjected to transfer learning,'' \emph{IEEE Transactions on Instrumentation and Measurement}, vol.~70, pp. 1--11, 2021.

\bibitem{doi:10.1177/1748006X20964614}
\BIBentryALTinterwordspacing
F.~Chen, L.~Liu, B.~Tang, B.~Chen, W.~Xiao, and F.~Zhang, ``A novel fusion approach of deep convolution neural network with auto-encoder and its application in planetary gearbox fault diagnosis,'' \emph{Proceedings of the Institution of Mechanical Engineers, Part O: Journal of Risk and Reliability}, vol. 235, no.~1, pp. 3--16, 2021. [Online]. Available: \url{https://doi.org/10.1177/1748006X20964614}
\BIBentrySTDinterwordspacing

\bibitem{7961149}
R.~Zhang, H.~Tao, L.~Wu, and Y.~Guan, ``Transfer learning with neural networks for bearing fault diagnosis in changing working conditions,'' \emph{IEEE Access}, vol.~5, pp. 14\,347--14\,357, 2017.

\bibitem{Zhou2020MultisignalVN}
\BIBentryALTinterwordspacing
J.~Zhou, X.~Yang, L.~Zhang, S.~Shao, and G.~Bian, ``Multisignal vgg19 network with transposed convolution for rotating machinery fault diagnosis based on deep transfer learning,'' \emph{Shock and Vibration}, vol. 2020, pp. 1--12, 2020. [Online]. Available: \url{https://api.semanticscholar.org/CorpusID:230549348}
\BIBentrySTDinterwordspacing

\bibitem{Jiang_2022}
\BIBentryALTinterwordspacing
L.~Jiang, C.~Zheng, and Y.~Li, ``Rotating machinery fault diagnosis based on transfer learning and an improved convolutional neural network,'' \emph{Measurement Science and Technology}, vol.~33, no.~10, p. 105012, jul 2022. [Online]. Available: \url{https://dx.doi.org/10.1088/1361-6501/ac7d3d}
\BIBentrySTDinterwordspacing

\bibitem{wang2021bearing}
X.~Wang, D.~Mao, and X.~Li, ``Bearing fault diagnosis based on vibro-acoustic data fusion and 1d-cnn network,'' \emph{Measurement}, vol. 173, p. 108518, 2021.

\bibitem{amarnath2013exploiting}
M.~Amarnath, V.~Sugumaran, and H.~Kumar, ``Exploiting sound signals for fault diagnosis of bearings using decision tree,'' \emph{Measurement}, vol.~46, no.~3, pp. 1250--1256, 2013.

\bibitem{liu2022anomalous}
Y.~Liu, J.~Guan, Q.~Zhu, and W.~Wang, ``Anomalous sound detection using spectral-temporal information fusion,'' in \emph{ICASSP 2022-2022 IEEE International Conference on Acoustics, Speech and Signal Processing (ICASSP)}.\hskip 1em plus 0.5em minus 0.4em\relax IEEE, 2022, pp. 816--820.

\bibitem{zhang2023anomalous}
H.~Zhang, J.~Guan, Q.~Zhu, F.~Xiao, and Y.~Liu, ``Anomalous sound detection using self-attention-based frequency pattern analysis of machine sounds,'' \emph{arXiv preprint arXiv:2308.14063}, 2023.

\bibitem{deng2019arcface}
J.~Deng, J.~Guo, N.~Xue, and S.~Zafeiriou, ``Arcface: Additive angular margin loss for deep face recognition,'' in \emph{Proceedings of the IEEE/CVF conference on computer vision and pattern recognition}, 2019, pp. 4690--4699.

\bibitem{anbalagan2023foundational}
S.~Anbalagan, D.~Agarwal, B.~Natarajan, and B.~Srinivasan, ``Foundational models for fault diagnosis of electrical motors,'' \emph{arXiv preprint arXiv:2307.16891}, 2023.

\bibitem{sehri2023university}
M.~Sehri, P.~Dumond, and M.~Bouchard, ``University of ottawa constant load and speed rolling-element bearing vibration and acoustic fault signature datasets,'' \emph{Data in Brief}, p. 109327, 2023.

\end{thebibliography}

\end{document}